\documentclass[10pt,conference]{IEEEtran}
\usepackage[letterpaper,margin=1in]{geometry}
\IEEEoverridecommandlockouts
\usepackage{cite}
\usepackage{amsmath,amssymb,amsfonts}
\usepackage{algorithmic}
\usepackage{graphicx}
\usepackage{textcomp}
\usepackage{xcolor}
\usepackage{stfloats}
\usepackage{tabularx}
\usepackage{booktabs}
\usepackage{multirow}
\def\BibTeX{{\rm B\kern-.05em{\sc i\kern-.025em b}\kern-.08em
    T\kern-.1667em\lower.7ex\hbox{E}\kern-.125emX}}
\begin{document}

\title{
Scaling Up Analog CIM: A Simulation\\
Framework for Topology-Aware Mapping
}

\author{
    \IEEEauthorblockN{
        Yuxin Ren, Danella Zhao, Huanrui Yang
    }
    \IEEEauthorblockA{
        Department of Electrical and Computer Engineering\\
        The University of Arizona, Tucson, AZ, USA\\
        \{yuxinr, danellazhao, huanruiyang\}@arizona.edu
    }
}

\maketitle

\begin{abstract}

Analog compute-in-memory (CIM) enables energy-efficient AI inference by performing in-situ matrix multiplication.
The growing demand for deploying larger models motivates chiplet integration to expand CIM capacity.
However, simulation abstractions based on local operator costs and aggregate communication metrics obscure the impact of weight-mapping topology when CIM system scales up.
We develop a phase analysis that quantifies how strongly mapping topology affects system performance across hardware configurations and how mapping optimization translates into latency gains.
To evaluate and exploit this design space, we present Isobar, a simulation platform integrating CIM decompositions, heterogeneous execution, and resource modeling.
Validated against silicon measurements and cycle-accurate simulation, Isobar provides reliable cost estimates that enable effective mapping search across model and hardware configurations.
For Gemma-7B prefill, Isobar-guided mapping reduces communication time by 23.3\% and end-to-end latency by 18.6\% relative to the strongest baseline.
Isobar's system-level modeling show that analog CIM retains energy-efficiency potential relative to GPU when scaled to thousands of dies, and identify topology-aware mapping as a key design dimension for efficient execution.

\end{abstract}

\section{Introduction}
\label{sec:introduction}

Serving large language models (LLMs) requires substantial compute resources~\cite{Kwon2023PagedAttention}.
Analog compute-in-memory (CIM) based on resistive random-access memory (ReRAM) enables efficiency by performing matrix operations in place~\cite{ISAAC2016}.
These benefits make deploying LLMs on analog CIM a promising direction~\cite{Chen2025ALBERTAnalogAI}.
However, scaling analog CIM to LLMs requires accommodating billions of parameters that far beyond individual CIM crossbar capacity, together with the difficulty in supporting the attention computation~\cite{Yang2020ReTransformer}.

Early CIM studies explored multi-die integration to expand memory capacity~\cite{SIAM2021} inspired by digital chiplet accelerators~\cite{shao2019simba}.
Heterogeneous architectures combine CIM dies with digital compute units to support attention~\cite{jaiswal2024halo}.
While expanding system capabilities, these architectural extensions introduce large volumes of data traffic across the CIM system: partitioning large resident weight matrices across dies requires transfers of inputs and intermediate results; heterogeneous execution adds communication among CIM dies, digital units, and memory resources.
Partitioning these matrices and assigning their blocks to physical CIM locations alone can produce different contention patterns on shared links and ports, leading to different communication delays even with the model and hardware configuration held fixed.
This assignment of resident computations to locations within a fixed hardware, which we call \emph{topology-aware mapping}, provides a software means of improving the efficiency.
Existing simulation methods do not fully capture this mapping dependence.
Digital chiplet frameworks emphasize pipelining and weight reloading~\cite{Gemini2024}, without modeling the decomposition and communication of resident analog execution.
Analog simulators focus on local operators or prescribed chiplet organizations~\cite{SIAM2021,2024_ispass_cimloop}, without resolving topology-dependent contention under alternative physical assignments.

\begin{figure*}[t]
    \centering
    \includegraphics[width=\textwidth]{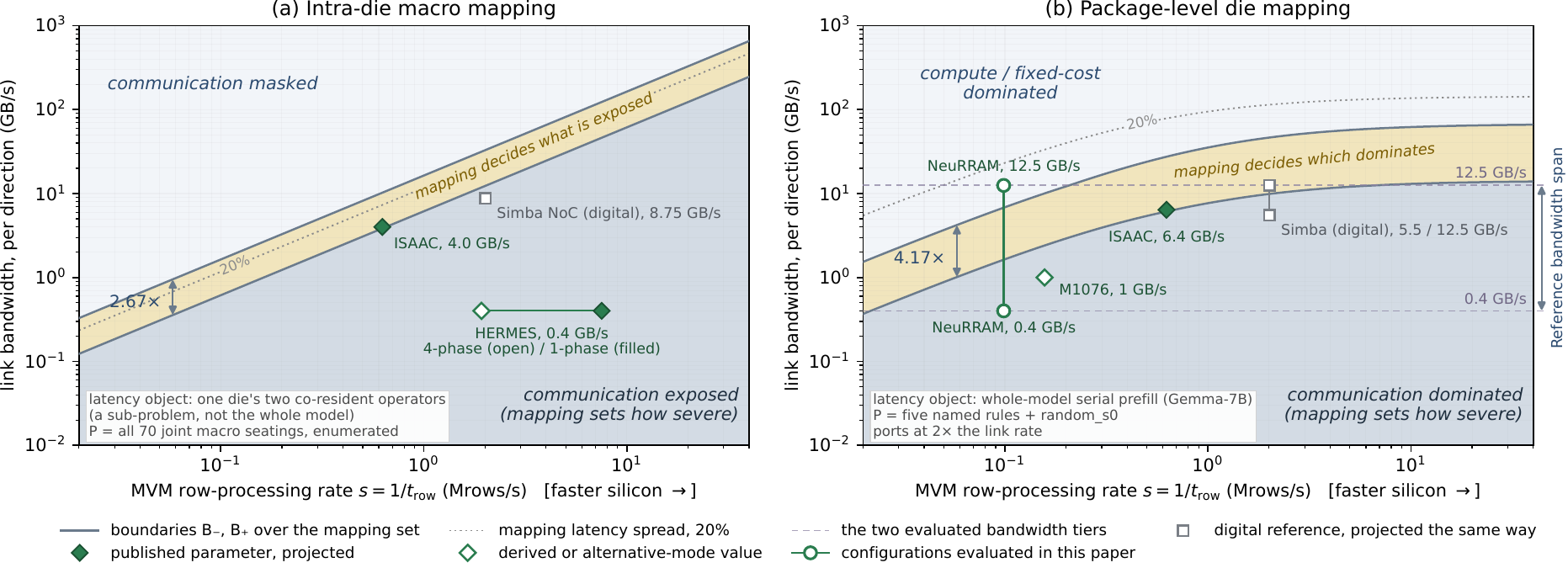}
    \caption{Mapping-dependent communication regimes for (a) intra-die macro mapping and (b) package-level die mapping.}
\label{fig:placement_phase}
\end{figure*}

To model and explore this space, we present Isobar, a simulation platform for topology-aware mapping in analog CIM chiplet systems.
Isobar derives communication from CIM decompositions and heterogeneous execution, analyzes the resulting contention on shared links and ports based on physical placement, and thereby constructs a latency and dynamic-energy model that connects feasible mapping choices to their consequences.
The resulting phase diagram (Fig.~\ref{fig:placement_phase}) quantitatively reveals the impact of topology-aware mapping, enabling comparison of deployments whose differences are hidden by previous metrics.
Across CIM speeds and interconnect bandwidths, the mapping-dependent boundary envelope identifies where physical assignment becomes critical as communication dominates intra-die or package-level execution.
Projected operating points from published CIM designs fall in mapping-sensitive regions under the reference workload, establishing the practicality of this design dimension.

Isobar further supports mapping search through incremental evaluation, enabling millisecond-scale evaluation and second-scale search with thousands of dies.
Comparisons with silicon measurements~\cite{shao2019simba} and independent BookSim simulations~\cite{Jiang2013BookSim} validate Isobar's ability to reliably search and evaluate different mappings.
Across three representative LLMs and three configurations, Isobar-guided mapping reduces median communication time by 14.8--24.0\% and end-to-end latency by 10.6--18.7\% relative to the strongest evaluated baselines.

We further show that analog CIM retains energy-efficiency potential at LLMs scale, with full-request dynamic-energy estimates well below published GPU measurements.
Controlled comparisons across capacity, matrix decomposition, and resource configurations reveal how these choices shape communication costs and mapping benefits, providing design guidance for efficient CIM scaling.

\section{Topology-Aware Mapping for Scaling Analog CIM}
\label{sec:topology_mapping}

\subsection{System Setting and Mapping Space}
\label{sec:system_mapping}

CIM deployment requires jointly determining where computation is allocated and how inputs and intermediate results are transferred, subject to hardware capacity constraints.
We formulate this mapping space according to the CIM hardware hierarchy.

Consider a CIM chiplet system with four hardware levels: package, die, macro, and crossbar array.
At the package level, we assume homogeneous CIM dies connected by a two-dimensional mesh.
Digital units and associated memory resources support input-dependent attention and KV-cache access, following previous CIM-based Transformer systems~\cite{jaiswal2024halo,Chen2025ALBERTAnalogAI}.
Within each die, an intra-die interconnect links multiple macros.
Each macro combines ReRAM crossbar arrays for static-weight matrix operations with data-conversion peripherals, local buffers, and digital accumulation logic~\cite{ISAAC2016}.
Array dimensions, macro and die capacities, interconnect geometry and bandwidth, and digital-resource organization are configurable in Isobar.

For CIM operators, weight precision, cell encoding, array dimensions, and die capacity constrain the legal matrix decompositions.
The weights and computation allocated to one macro form a \emph{logical macro instance}; the macro instances allocated to one die form a \emph{logical die instance}.
We represent each candidate deployment as a triple:
\begin{equation}
p=(\gamma,\alpha,\beta).
\label{eq:deployment_triple}
\end{equation}
Here, $\gamma=((G_1,\mathcal{A}_1),\ldots,(G_K,\mathcal{A}_K))$ partitions the model's CIM-mapped linear operators into nonempty, consecutive groups $G_k$.
Each $\mathcal{A}_k$ specifies how the operators' weight matrices are legally tiled and how the resulting encoded weight tiles are allocated to logical macro and die instances.
The tilings also determine which tiles share inputs and which partial results must be summed~\cite{2024_ispass_cimloop}.
The resulting logical die instances form $\mathcal{D}_{\gamma}$, and $\mathcal{M}_d$ denotes the macro instances belonging to logical die $d$.

Given $\gamma$, let $\mathcal{S}$ denote the available CIM die slots in the package and $\mathcal{Q}_s$ the macro slots within physical die $s$.
Die placement $\alpha$ assigns each logical die to a package slot, while macro assignment $\beta_d$ assigns its macro instances to slots within the selected die $\alpha(d)$:
\begin{equation}
\alpha:\mathcal{D}_{\gamma}\rightarrow\mathcal{S},\qquad
\beta_d:\mathcal{M}_d\rightarrow\mathcal{Q}_{\alpha(d)},\quad d\in\mathcal{D}_{\gamma},
\label{eq:physical_assignments}
\end{equation}
with $\beta=\{\beta_d\}_{d\in\mathcal{D}_{\gamma}}$ collecting the intra-die assignments.
Both $\alpha$ and each $\beta_d$ are injective, so each physical slot hosts at most one corresponding resident instance; unused slots are permitted.
The feasible deployment space $\mathcal{F}$ consists of triples that respect the grouping and slot-exclusivity constraints and satisfy each instance's capacity and resource-type requirements.

For placement comparisons, we fix $\gamma$, the workload, hardware, routing policy, and execution policy, including digital-unit locations and task allocation.
Varying $\alpha$ or $\beta$ changes transfer routes and contention on shared links and ports while preserving operator computations and logical data transfers.

\subsection{Execution Costs and Boundary Envelopes}
\label{sec:execution_boundaries}

Under the mapping space in Eq.~\eqref{eq:deployment_triple}, we combine communication, computation, and memory-access costs to quantify how topology-aware mapping affects latency across hardware settings.

For inter-die placement comparisons, non-communication cost is
\begin{equation}
R(s)=R_{\mathrm{CIM}}(s)+R_{\mathrm{digital}}+R_{\mathrm{KV}}+R_{\mathrm{other}},
\label{eq:noncommunication_cost}
\end{equation}
where $s=1/t_{\mathrm{row}}$ is the CIM row-processing rate, and the four terms represent CIM execution, digital computation, KV-memory service, and other processing, respectively.
Communication cost depends on how mapping distributes traffic across shared resources.
For a stage $j$ of simultaneously ready transfers, let $V_{j,r}(p)$ denote the total bytes routed through resource $r$ under mapping $p$.
The resource set $\mathcal{R}_j$ includes directed links, injection and ejection ports, and digital-unit interfaces used by that stage.
The communication-time estimate is
\begin{equation}
H(p;\mathbf{C})=
\sum_j\left[P_j(p)+
\max_{r\in\mathcal{R}_j}\frac{V_{j,r}(p)}{C_r}\right],
\label{eq:resource_communication}
\end{equation}
where $C_r$ is the effective bandwidth of $r$, $\mathbf{C}$ collects these bandwidths, and $P_j(p)$ is the stage's propagation delay.
The maximum term is a lower bound on the stage's service time, used as its estimate under the reference static model.
Package communication and local execution are serialized under the reference policy, giving
\begin{equation}
T_{\mathrm{pkg}}(p;s,\mathbf{C})=R(s)+H(p;\mathbf{C}).
\label{eq:package_latency}
\end{equation}
In addition, fixed digital-interface bandwidths must be accounted when the mesh bandwidth $B$ varies:
\begin{equation}
H(p;B,\kappa)=
\sum_j\left[P_j(p)+
\max\left\{\frac{A_j(p;\kappa)}{B},U_j(p)\right\}\right],
\label{eq:independent_interface_cost}
\end{equation}
Where $\kappa$ refers to CIM input and output ports bandwidth, $A_j$ is the maximum over mesh-link byte loads and CIM-port byte loads, $U_j$ is the largest service time among the fixed-bandwidth interfaces used by stage $j$.
Once a fixed-bandwidth interface becomes the bottleneck, further increases in $B$ cannot shorten that stage.

The \emph{inter-die boundary} marks where communication time falls to or below the non-communication cost in the serialized package execution.
For fixed $s$, $\kappa$, and digital-interface bandwidths, we define
\begin{equation}
B_c(p;s)=
\inf\left\{B>0:H(p;B,\kappa)\le R(s)\right\}.
\label{eq:package_boundary}
\end{equation}
This boundary separates communication-dominated execution, $H>R(s)$, from execution where non-communication work contributes at least as much latency, $H\le R(s)$.
Communication remains additive on both sides.
An empty set gives $B_c=+\infty$.

Within a die, the reference schedule permits computation and communication service to overlap:
\begin{equation}
T_{\mathrm{die}}(\beta;s,B)=
\sum_j\left[D_j(\beta)+
\max\{C_j(s),H_j(\beta;B)\}\right],
\label{eq:intradie_latency}
\end{equation}
where $C_j(s)$ and $H_j(\beta;B)$ are $j$'s computation and communication service times, respectively, and $D_j(\beta)$ includes non-overlapped costs such as writes and propagation.
The \emph{intra-die boundary} therefore marks where communication service can be hidden by computation for every co-resident operator:
\begin{equation}
b(\beta;s)=
\inf\left\{B>0:H_j(\beta;B)\le C_j(s)\ \forall j\right\}.
\label{eq:intradie_boundary}
\end{equation}

Different mappings reach these boundaries at different bandwidths.
For package mappings $\mathcal{P}\subseteq\mathcal{F}$ evaluated under the same fixed conditions, the lowest and highest thresholds define the observed boundary envelope:
\begin{equation}
B_-(s)=\min_{p\in\mathcal{P}}B_c(p;s),
\qquad
B_+(s)=\max_{p\in\mathcal{P}}B_c(p;s).
\label{eq:boundary_envelope}
\end{equation}
The intra-die envelope is obtained analogously from the extrema of $b(\beta;s)$ over the evaluated macro assignments.
At each CIM rate, the envelope divides bandwidth into three regions.
Below $B_-(s)$, every evaluated mapping has communication-dominated execution at package level or exposed communication service at die level.
Above $B_+(s)$, every evaluated mapping satisfies $H\le R(s)$ at package level or hides communication service for every operator at die level.
Within the envelope, mapping determines which side of the corresponding execution boundary is reached.
$B_+/B_-$ measures the variation in required bandwidth across the evaluated mappings.

\begin{figure*}[t]
    \centering
    \includegraphics[width=\textwidth]{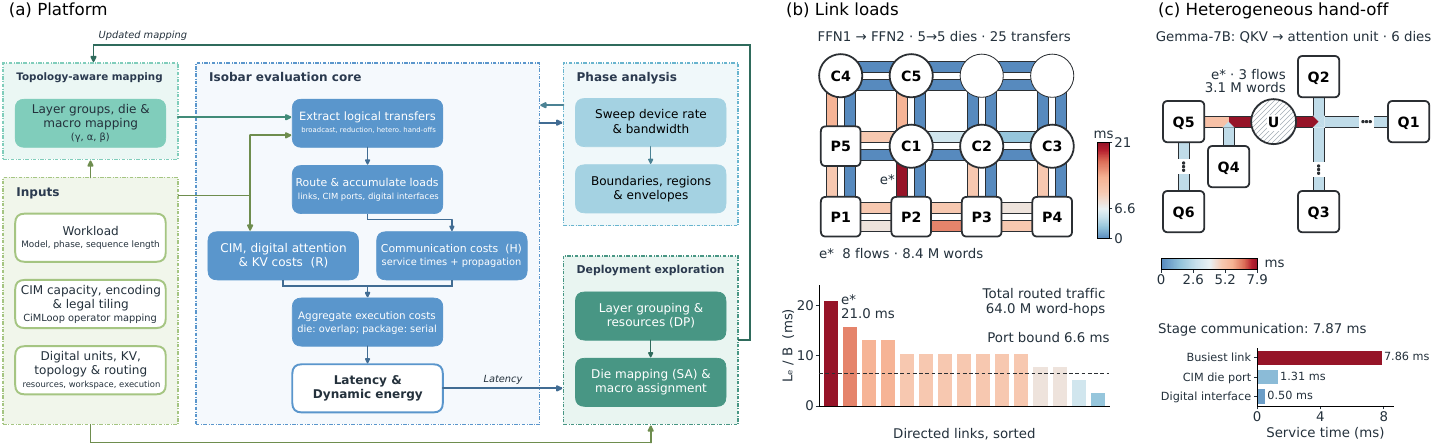}

    \caption{Isobar framework and communication modeling: (a) simulation workflow supporting phase analysis and mapping search; (b) shared-link contention in CIM--CIM transfers; (c) link, CIM-port, and digital-interface service times for a CIM--digital attention handoff.}
    \label{fig:isobar_overview}
\end{figure*}

\subsection{Phase Interpretation and Design Implications}
\label{sec:phase_implications}

Using the preceding models, we sweep the CIM processing rate and interconnect bandwidth to quantify how mapping topology affects the boundary bandwidth and total latency, as shown in Fig.~\ref{fig:placement_phase}.
At the reference speed, the ratio between the largest and smallest boundary bandwidths is 2.67$\times$ across 70 intra-die macro assignments and 4.17$\times$ across six Gemma-7B inter-die mappings.
Thus, different mapping topologies require substantially different bandwidths to reach the same boundary.
The dotted contours mark a total-latency spread of $(T_{\max}-T_{\min})/T_{\min}=20\%$ across the evaluated mappings, indicating a 20\% difference attributable to mapping topology alone under the same workload, CIM speed, and interconnect bandwidth.

We further project existing CIM designs onto the phase diagrams to relate this variation to real hardware capabilities.
ISAAC~\cite{ISAAC2016} provides an architectural reference for ReRAM acceleration.
HERMES~\cite{LeGallo2023Hermes} and NeuRRAM~\cite{Wan2022NeuRRAM} provide silicon implementations of PCM- and RRAM-based CIM, respectively, while M1076~\cite{Mythic2021M1076} represents a flash-based analog matrix processor.
The reference configurations span a range of analog CIM baselines, from slow to fast, with different communication capabilities.
Both HERMES operating modes are included to represent its speed--precision tradeoff between one-phase and four-phase computation.
We also include Simba~\cite{zimmer2020scalable}, a digital multi-chiplet accelerator, as a reference for ideal communication provisioning at the package level, using its range of demonstrated digital chiplet bandwidths.
For the NeuRRAM-derived configurations, we pair NeuRRAM's compute timing with two bandwidth settings: HERMES's 0.4~GB/s on-chip link rate and Simba's maximum 12.5~GB/s die-to-die link rate.
These endpoints define a range from an analog-CIM interconnect reference to an optimistic digital-chiplet reference, bracketing the boundary at the reference CIM speed.

Under the reference workloads, most projected design points lie in regions sensitive to topology-aware mapping, establishing it as a key determinant of execution efficiency for CIM systems with existing compute and interconnect capabilities.
Capacity expansion also changes this sensitivity: when resident matrices must span more dies, input broadcasts, partial-sum reductions, and operator handoffs alter shared-resource loads and contention, shifting the phase boundaries.
By deriving these loads from legal CIM decompositions, Isobar quantifies the communication cost of capacity expansion and enables the evaluation and optimization of topology-aware mappings.

\section{The Isobar Platform}
\label{sec:isobar}

\subsection{Simulation Framework}
\label{sec:modeling_approach}

Figure~\ref{fig:isobar_overview}(a) summarizes Isobar's simulation workflow, which takes a workload description, hardware and execution settings, and a candidate deployment as input.
The evaluation core derives communication demands and component costs, then returns total latency, dynamic energy, and resource-level cost breakdowns.

Starting from CiMLoop's decompositions~\cite{2024_ispass_cimloop}, Isobar uses tensor indices and reuse relationships to identify shared inputs and contributions to outputs, while operator dependencies determine producer--consumer handoffs.
These relationships are converted into broadcasts, reductions, and handoffs, with each transfer's payload, endpoints, and execution stage recorded.
Payloads are computed from the mapped tensor slices and their activation or accumulation, preserving the communication requirements of the chosen matrix decomposition.

Isobar supports prefill and decode with either a central digital attention unit or multiple distributed units.
Unit locations, compute throughput, memory resources, and workspace capacities are configurable, and tasks can be assigned by layer or head group.
These assignments determine which QKV outputs or partial sums each unit receives and where its outputs return to the CIM output projection.
KV reads and writes follow the configured storage paths and are accounted for separately from transfers between compute units.

Isobar routes transfers under each candidate deployment to obtain the resource loads in Eq.~\eqref{eq:resource_communication}.
Figure~\ref{fig:isobar_overview}(b) illustrates this accumulation: eight FFN handoff flows share the busiest link, whose 21.0~ms service time exceeds the 6.6~ms port bound, making shared-link contention the communication bottleneck.
The heterogeneous handoff (Fig.~\ref{fig:isobar_overview}(c)) is also limited by contention: the QKV-to-attention's 7.86~ms service time exceeds those of both the CIM port and the digital interface and determines the stage's 7.87~ms communication latency.

Isobar combines the resulting communication costs with CIM operator costs, digital computation, and memory service.
Digital computation costs follow operation counts and the effective throughput of the assigned units.
The memory model supports both materialized and tiled attention, accounting for KV reads and writes, intermediate spills, and block rereads under the configured workspace capacity.
Automatic selection chooses the attention mode with lower modeled memory traffic.
These costs are aggregated along the execution graph using the overlap and serialization rules in Section~\ref{sec:execution_boundaries}.
When multiple digital units cooperate on an attention stage, stage completion follows their critical path.

Dynamic energy is accumulated over the same execution:
\begin{equation}
E_{\mathrm{dyn}}=
E_{\mathrm{CIM}}+E_{\mathrm{digital}}+
E_{\mathrm{KV}}+E_{\mathrm{net}}+E_{\mathrm{other}}.
\label{eq:dynamic_energy}
\end{equation}
The non-network terms follow the corresponding operator and memory-access costs.
Network energy is computed as $E_{\mathrm{net}}(p)=\sum_{\ell}e_{\ell}V_{\ell}(p)$, where $V_{\ell}(p)$ is the cumulative byte volume on link $\ell$ and $e_{\ell}$ is its energy per byte.
This accounting retains the different effects of resource contention on latency and cumulative transmission work on energy.
The shared evaluation core supports both the phase analysis and topology-aware mapping optimization.

\begin{figure}[t]
    \centering
    \includegraphics[width=0.98\columnwidth]{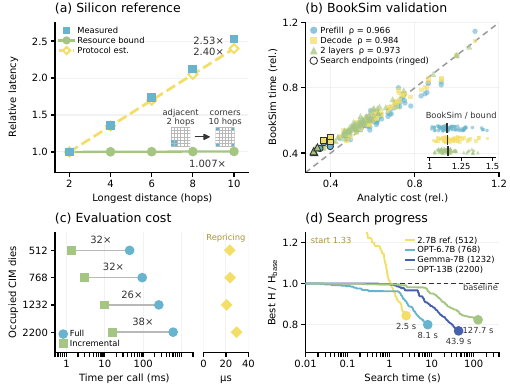}

    \caption{Isobar accuracy and efficiency: (a) latency vs. chiplet separation validation; (b) latency and ranking agreement with BookSim; (c) full evaluation, incremental update, and repricing runtimes; (d) comm. cost vs. search time.}
\label{fig:external_validation}
\end{figure}

\subsection{Efficient Exploration and Validation}
\label{sec:validation}

To use the framework for topology-aware mapping optimization, Isobar implements a hierarchical dynamic-programming--simulated-annealing (DP--SA) search over the deployment choices defined in Section~\ref{sec:system_mapping}.
Candidate deployments are scored using simulated execution latency $T(\gamma,\alpha,\beta)$.
DP selects the consecutive groups $G_k$ of CIM-mapped linear operators, with their allocations $\mathcal{A}_k$ determined by the fixed operator tilings and allocation rules $\gamma$.
SA then optimizes the physical die assignment $\alpha$.
For small intra-die assignment spaces, Isobar exhaustively enumerates the joint macro assignments $\beta$.
SA explores die exchanges, relocations, and group swaps that preserve allocation sizes and slot exclusivity.
Improving moves are accepted directly, while uphill moves are accepted probabilistically under geometric cooling.
Multiple starts explore alternative initial assignments, and each run retains its best visited mapping.
The main inter-die experiments fix $\gamma$ and $\beta$ and use $H/H_{\mathrm{base}}$ as the SA objective, where $H_{\mathrm{base}}$ is the communication time of a fixed baseline.

Fig.~\ref{fig:external_validation} validates Isobar's predictions.
Simba's silicon measurement test runs the same workload on increasingly separated chiplets, directly measuring the latency penalty of physical placement~\cite{shao2019simba}.
Isobar reproduces the measured latency increase with chiplet separation, with only 5.1\% relative error in the latency ratio between the most separated and adjacent placements, while methods that omit the transfer schedule fail to capture this trend (Fig.~\ref{fig:external_validation}(a)).
We further validate the communication model against BookSim~\cite{Jiang2013BookSim}, a cycle-accurate network simulator that provides a high-fidelity external reference for latency.
Under different configurations, Isobar's mapping rankings keep closely agree with BookSim's (Fig.~\ref{fig:external_validation}(b)).

To accelerate repeated evaluation, Isobar precomputes logical transfers and placement-independent execution costs for each fixed $\gamma$.
Each SA move updates only the affected routes, resource loads, and stage costs.
Parameter sweeps also reuse the decomposition when logical traffic and feasibility remain unchanged.
Across deployments with 512--2,200 CIM dies, incremental evaluation takes 1.4--15.9\,ms, achieving a 26--38$\times$ speedup over full evaluation (Fig.~\ref{fig:external_validation}(c)). The representative search runs in Fig.~\ref{fig:external_validation}(d) finish in 2.5--127.7\,s, each producing a mapping with lower communication cost than its baseline.
Together, these results demonstrate that Isobar retains the predictive accuracy needed to guide mapping decisions while making repeated evaluation practical at thousands-of-dies scale.

\section{Mapping Benefits and Scaling Behavior}
\label{sec:deployment}

\begin{table}[t]
    \caption{Reference settings and evaluation protocols.}
    \label{tab:setup}
    \centering
    \setlength{\tabcolsep}{3pt}
    \renewcommand{\arraystretch}{1.04}
    \begin{tabularx}{0.98\columnwidth}{
        @{}>{\raggedright\arraybackslash}p{0.34\columnwidth}
        >{\raggedright\arraybackslash}X@{}
    }
        \toprule
        \textbf{Parameter} & \textbf{Setting} \\
        \midrule
        OPT-6.7B deployment
            & 768 CIM dies; $28\times28$ mesh \\
        Gemma-7B deployment
            & 1,232 CIM dies; $36\times36$ mesh \\
        OPT-13B deployment
            & 2,200 CIM dies; $47\times47$ mesh \\
        Input
            & Batch 1; 512-token prefill \\
        \midrule
        CIM die
            & 8 macros; 256 arrays of $256\times256$ cells \\
        Cell / encoding
            & 4 bits/cell; differential weights \\
        Activations / sums
            & 8-bit activations; 16-bit partial sums \\
        CIM row time
            & $10.14\,\mu\mathrm{s}$, NeuRRAM-derived~\cite{Wan2022NeuRRAM} \\
        Digital attention
            & 1 central unit; 2 TMAC/s effective \\
        KV memory
            & 2 GB; 25.6 GB/s \\
        Workspace
            & 8 MB \\
        Digital interface
            & 12.5 GB/s in each direction \\
        Network
            & 2D mesh; XY routing; 20 ns/hop \\
        Link / CIM port
            & 0.4 / 0.8 GB/s per direction ($\kappa=2$) \\
        Network energy
            & 10 pJ per 8-bit word-hop \\
        Digital energy
            & 1 pJ per MAC-equivalent operation \\
        \midrule
        Main baseline pool
            & Feasible prescribed rules + 500 random layouts \\
        Main search runs
            & 10 seeds; 3 starts per seed \\
        Exploratory sweeps
            & 100 random layouts; 5 seeds \\
      \bottomrule
    \end{tabularx}
\end{table}

\subsection{Evaluation Setup}
\label{sec:evaluation_setup}

Table~\ref{tab:setup} lists the reference settings, with Gemma-7B as the main case.
The models are also evaluated under two variants: 16 macros per CIM die, or four digital attention units assigned separate head groups while retaining the reference total digital-resource budget.
The mapping experiments report Transformer-backbone latency and dynamic energy.
For each configuration, the strongest baseline is the lowest-latency candidate in the pool.

\begin{figure}[t]
    \centering
    \includegraphics[width=\columnwidth]{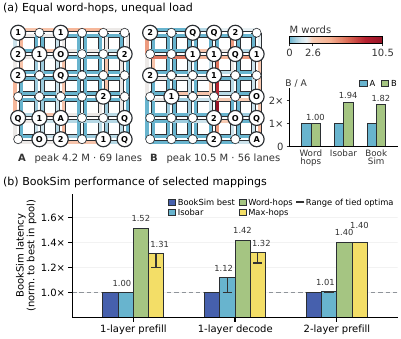}
    \caption{Mapping discrimination and selection quality: (a) equal word-hops yield different link loads and latencies; (b) BookSim latency of each metric's selected mapping, normalized to the best candidate in the same pool.}
\label{fig:mapping_discriminability}
\end{figure}

\begin{figure}[t]
    \centering
    \includegraphics[width=\columnwidth]{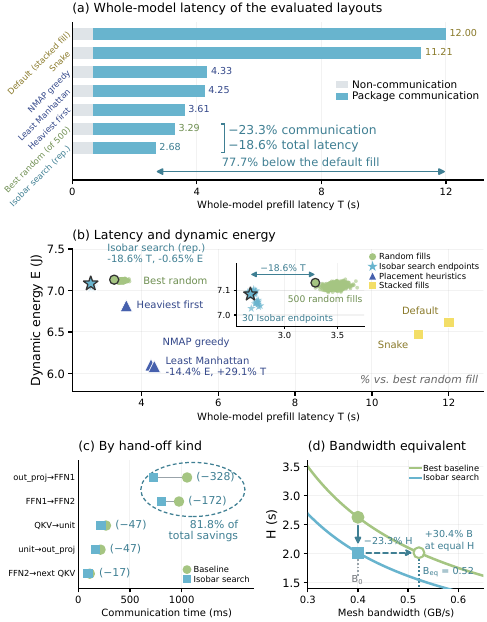}
    \caption{Topology-aware mapping benefits for Gemma-7B prefill: (a) backbone latency across mappings; (b) latency--dynamic-energy trade-offs; (c) communication time and savings by handoff, summed over layers; (d) bandwidth required by the strongest baseline to match the searched mapping at $0.4$~GB/s.}
    \label{fig:communication_benefits}
\end{figure}

\subsection{Mapping Discrimination and Selection Quality}
\label{sec:discriminability}

Figure~\ref{fig:mapping_discriminability}(a) compares two mappings with identical logical transfers but different traffic distributions.
Mapping B concentrates traffic on fewer links, increasing the peak link load from 4.2~M words in A to 10.5~M words.
The Word-hops metric, which sums transferred words over all traversed links, assigns the two mappings equal costs.
Isobar distinguishes their contention patterns and predicts a B/A communication-time ratio of 1.94, similar to 1.82 in BookSim.

Figure~\ref{fig:mapping_discriminability}(b) evaluates whether this discrimination improves mapping selection in three different settings, using the candidate pools from Section~\ref{sec:validation}.
For each workload, Isobar, Word-hops, and Max-hops each select the candidate with the lowest predicted cost; Max-hops uses the maximum route length.
The selected mappings are then evaluated in BookSim, with their communication times normalized to the minimum BookSim time within the same pool.
Thus, the bars compare the quality of mapping choices made by the three metrics.

Isobar selects mappings whose BookSim communication times are only 0.3--12.2\% above the pool minimum, compared with 40.4--51.7\% for Word-hops, and outperforms Max-hops in all three workloads.
The range bars show the BookSim variation among candidates tied for each metric's minimum predicted cost.
For single-layer decode, Isobar's tied candidates include the BookSim-best mapping, while the fixed tie-breaking rule selects the mapping with 12.2\% higher latency.
These results demonstrate that accounting for shared-resource contention improves topology-aware mapping selection.

\begin{figure}[t]
    \centering
    \includegraphics[width=\columnwidth]{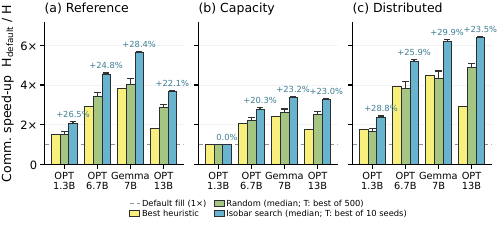}
    \caption{Communication speedups over default fill: (a) 8 macros/die and 1 digital unit; (b) 16 macros/die and 1 unit; (c) 8 macros/die and 4 head-group units. Bars show medians; caps mark the best. Percentages indicate communication-time reductions relative to the random median.}
    \label{fig:scaling_benefits}
\end{figure}

\begin{figure}[t]
    \centering
    \includegraphics[width=0.48\textwidth]{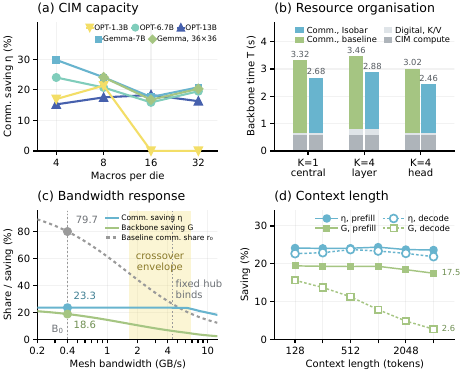}
    \caption{Configuration dependence of mapping gains: (a) CIM capacity with a fixed-mesh; (b) digital-unit organization at fixed resources; (c) bandwidth response for a fixed mapping; (d) comm. and backbone latency reductions vs. context length in prefill and decode. Panels (b)--(d) use Gemma-7B.}
    \label{fig:configuration_dependence}
\end{figure}

\subsection{Communication Benefits and Bandwidth Implications}
\label{sec:communication_benefits}

We demonstrate Isobar's optimization performance across architectural configurations.
For Gemma-7B prefill, Fig.~\ref{fig:communication_benefits}(a) shows that Isobar reduces communication time by 77.7\% to the default mapping, and 23.3\% to the strongest baseline, indicating a 18.6\% end-to-end improvement.
Figure~\ref{fig:communication_benefits}(b) shows that this latency improvement is accompanied by a 0.65\% reduction in dynamic energy relative to the same baseline.
Least Manhattan achieves a larger energy reduction of 14.4\% but increases latency by 29.1\%, illustrating the different mapping choices favored by latency and energy objectives.

Figure~\ref{fig:communication_benefits}(c) locates the main source of the communication savings.
Transfers from the attention output to FFN1 and from FFN1 to FFN2 together account for 81.8\% of the reduction, indicating shared links as the bottlenecks in the baseline handoffs, linking these gains to reduced contention along the FFN communication paths.
Figure~\ref{fig:communication_benefits}(d) shows that topology-aware mapping can reduce the interconnect bandwidth needed to meet a communication target.
With the digital-interface capacity fixed, the baseline requires 30.4\% more bandwidth to match the searched mapping's communication time.

Figure~\ref{fig:scaling_benefits} extends the comparison across model scales and attention configurations, improving the strongest baselines across all three configurations for the larger models.
The smaller OPT-1.3B case shows that these gains depend on the communication structure produced by configuration, as capacity consolidation of certain settings leaves little contention to reduce.
The results establish Isobar's ability to optimize topology-aware mappings across architectures.

\begin{table}[t]
    \centering
    \caption{QKV partitioning for OPT-13B prefill
    ($S=512$, $49\times49$ mesh).}
    \label{tab:qkv_partition}
    \setlength{\tabcolsep}{2pt}
    \renewcommand{\arraystretch}{1.04}

    \begin{tabular*}{0.9\columnwidth}
        {@{\extracolsep{\fill}}lccccc@{}}
        \toprule
        \shortstack{QKV\\split} &
        \shortstack{CIM\\dies} &
        \shortstack{QKV-to-unit\\traffic (GB)} &
        \shortstack{$T_{\mathrm{base}}$\\(s)} &
        \shortstack{$T_{\mathrm{search}}$\\(s)} &
        \shortstack{$\eta$\\(\%)} \\
        \midrule
        Row    & 2200 & 6.29  & 13.39 & 11.22 & 17.5 \\
        Column & 2400 & 0.315 &  5.16 &  4.10 & 25.1 \\
        \bottomrule
    \end{tabular*}

\end{table}

\subsection{Configuration Dependence and Design Guidance}
\label{sec:configuration_dependence}

Isobar's flexible configuration supports design studies that guide capacity provisioning, resource organization across workloads.

First, capacity choices should account for the communication patterns produced by matrix decompositions.
Figure~\ref{fig:configuration_dependence}(a) shows that increasing die capacity does not consistently reduce the benefits of topology-aware mapping.
Capacity consolidation removes competing handoff flows in OPT-1.3B, whereas wide fan-out handoffs retain concentrated contention in Gemma-7B.
The Gemma-7B common-grid control preserves the nonmonotonic trend, showing that mesh size alone do not explain this behavior.
Table~\ref{tab:qkv_partition} further illustrates the trade-off between resident footprint and communication demand.
Column partitioning uses more dies but replaces repeated partial-sum transfers with disjoint QKV outputs, reducing QKV-to-unit traffic by $\sim$20$\times$ and substantially lowering latency.
Capacity, decomposition, and topology-aware mapping should therefore be evaluated jointly, since minimizing one does not necessarily minimize total cost.

Second, resource distribution must preserve computational parallelism while balancing communication and memory.
Figure~\ref{fig:configuration_dependence}(b) shows that assigning attention head groups to separate units disperses communication while preserving the aggregate throughput.
Assigning complete layers to separate units instead leaves part of the digital capacity idle, increasing latency relative to the central-unit configuration.
Workspace allocation introduces another constraint: at longer contexts, dividing the workspace among distributed units requires more query tiles and repeated KV reads, offsetting part of the communication benefit.
Complementary port sweeps show that tight CIM-port limits also restrict mapping gains.

Third, the system benefit of communication optimization depends on the execution costs.
Figure~\ref{fig:configuration_dependence}(c) illustrates this dependence by sweeping mesh bandwidth for a fixed mapping pair.
Total latency savings initially decline while fractional communication savings remain nearly constant, as communication accounts for a smaller share of execution.
Once the fixed-bandwidth digital interface becomes limiting, communication savings also decline.
Figure~\ref{fig:configuration_dependence}(d) shows a related effect across workloads.
As the decode context grows, topology-aware mapping retains similar communication savings, but increasing KV service costs progressively reduce their contribution to backbone acceleration.
Prefill retains larger and more stable backbone gains across the evaluated context lengths.
Long-context decode therefore requires attention to memory service alongside communication optimization, whereas prefill offers greater scope for acceleration through topology-aware mapping.

By exposing these dependencies, Isobar supports the joint evaluation and optimization of hardware configurations and topology-aware mappings according to the dominant costs of each deployment.

\begin{table}[t]
    \centering
    \caption{Reported GPU and modeled CIM energy per inference request.}
    \label{tab:gpu-energy}
    \setlength{\tabcolsep}{3pt}
    \renewcommand{\arraystretch}{1.04}

    \begin{tabular*}{0.9\columnwidth}
        {@{\extracolsep{\fill}}ccrrc@{}}
        \toprule
        Model & Macros/die & GPU (J) & CIM (J) & GPU/CIM \\
        \midrule
        \multirow{2}{*}{OPT-13B}
        & 8  & \multirow{2}{*}{5,473.7} & 58.61 & $93\times$  \\
        & 16 &                         & 48.53 & $113\times$ \\
        \midrule
        \multirow{2}{*}{Gemma-7B}
        & 8  & \multirow{2}{*}{1,898.7} & 11.45 & $166\times$ \\
        & 16 &                         &  9.85 & $193\times$ \\
        \bottomrule
    \end{tabular*}

    \par\vspace{2pt}
    \begin{minipage}{\columnwidth}
        \scriptsize
        GPU energy from A100~\cite{cxlpnm2024} for OPT-13B, and average requests from Jetson AGX Thor board~\cite{hydra2026} for Gemma-7B. CIM values are simulated dynamic energy.
    \end{minipage}
\end{table}

\subsection{Energy Comparison with GPUs}
\label{sec:gpu-energy}

Using the reference hardware settings and the searched mapping for each capacity configuration, we estimate dynamic energy for complete inference requests, including prefill, decoding, embeddings, and additional resident LM-head resources.
At batch size one with 16-bit KV caches, we match the published GPU workloads: 64 input and 1,024 output tokens for OPT-13B, and 13--345 input and 500 output tokens for Gemma-7B.
Table~\ref{tab:gpu-energy} shows that modeled CIM dynamic energy remains substantially below reported GPU energy after accounting for communication and non-CIM execution.

\section{Related Work}
\label{sec:related_work}

Existing analog CIM simulators model analog nonidealities~\cite{crosssim}, architectures, and workload mappings~\cite{2024_ispass_cimloop}.~\cite{hu2005mapping} optimize NoC mapping and routing, NMAP~\cite{nmap} incorporates link-bandwidth constraints, and HALO~\cite{jaiswal2024halo} optimizes placement in heterogeneous IMC--CMOS systems using traffic-weighted distance, leaving mappings with equal distance but different bottleneck loads unchecked.

Timeloop~\cite{timeloop} supports analytical loop-mapping exploration, Gemini~\cite{Gemini2024} co-explores graph partitioning, spatial mapping, and chiplet architectures, and RapidChiplet~\cite{RapidChiplet} provides fast interconnect evaluation.
SIAM~\cite{SIAM2021} integrates computation, communication, and memory costs, and CHIPSIM~\cite{pfromm2025chipsim} models contention across concurrent workloads and pipelining.
These tools primarily target digital accelerators.

Isobar jointly models resident CIM capacity constraints, heterogeneous execution, and shared-resource contention to evaluate and optimize topology-aware mappings across configurations.

\section{Conclusion}
\label{sec:conclusion}

Isobar provides a configurable framework for evaluating and optimizing topology-aware mappings in large-scale analog CIM systems.
It connects legal CIM decompositions, heterogeneous execution, and shared-resource contention to capture performance differences that aggregate traffic metrics miss.
Validation against silicon measurements and BookSim supports its communication estimates and mapping selection, while incremental evaluation makes exploration practical at thousands-of-dies scale.
The phase and configuration studies establish the importance of topology-aware mapping under existing hardware capabilities and identify how capacity, resource organization, and workload govern its system benefits.
Full-request comparisons further demonstrate the energy-efficiency potential of these deployments, with modeled CIM dynamic energy remaining substantially below reported GPU energy after accounting for communication and non-CIM execution.
These results establish topology-aware mapping as an integral part of CIM architecture design, with Isobar supporting hardware configuration choices and mapping optimization through a shared evaluation framework.

\section*{Acknowledgment}
This work is supported in part by TetraMem, Inc.,
through a research collaboration grant and computational resources.

\clearpage

\bibliographystyle{IEEEtran}
\bibliography{reference}

\end{document}